\documentclass[11pt]{amsart}
\usepackage{amsmath}
\usepackage{amssymb}
\usepackage{amsthm}
\usepackage{fullpage}
\usepackage{xy}
\usepackage{url}

\long\def\symbolfootnote[#1]#2{\begingroup%
\def\thefootnote{\fnsymbol{footnote}}\footnote[#1]{#2}\endgroup}

\long\def\symbolfootnotemark[#1]{\begingroup%
\def\thefootnote{\fnsymbol{footnote}}\footnotemark[#1]\endgroup}

\theoremstyle{definition} 
\newtheorem{defn}{Definition}[section]
\newtheorem{open}[defn]{Open Question}
\newtheorem{example}[defn]{Example}

\theoremstyle{plain}
\newtheorem{prop}[defn]{Proposition}
\newtheorem{lem}[defn]{Lemma}
\newtheorem{cor}[defn]{Corollary}
\newtheorem{thm}[defn]{Theorem}

\input xy
\xyoption{all}

\newcommand{\ie}{i.\,e.}
\newcommand{\eg}{e.\,g.}

\newcommand{\definedWord}[1]{\emph{#1}}
\newcommand{\dash}{\text{-}}

\newcommand{\Z}{\ensuremath{\mathbb{Z}}}

\newcommand{\otherwise}{\text{ otherwise}}
\newcommand{\If}{\text{ if }}
\newcommand{\Iff}{\ensuremath{\Longleftrightarrow}}
\newcommand{\defeq}{\stackrel{def}{=}}

\newcommand{\ifftext}{if and only if }

\newcommand{\union}{\ensuremath{\cup}}
\newcommand{\set}[1]{\{#1\}}
\newcommand{\tuple}[1]{\langle #1 \rangle}

\DeclareMathOperator{\grph}{graph}

\DeclareMathOperator{\dom}{dom}

\DeclareMathOperator{\Ker}{Ker}
\newcommand{\cc}[1]{\ensuremath{\mathsf{#1}}}
\newcommand{\lang}[1]{\mathit{#1}}
\newcommand{\ccCF}[1]{\cc{CF(#1)}}

\newcommand{\ccKer}[1]{\cc{Ker(#1)}}
\newcommand{\ccLexEq}[1]{\cc{LexEq #1}}
\newcommand{\FCFshort}{\cc{LexEq}}
\newcommand{\CFshort}{\cc{CF}}
\newcommand{\Kershort}{\cc{Ker}}
\newcommand{\Eqshort}{\cc{PEq}}
\newcommand{\partialFP}{PF}

\newcommand{\oracle}[1]{^{\lang{#1}}}

\newcommand{\orig}{}

\newtheorem*{collisionprop}{Proposition~\ref{prop:collision-free}}
\newcommand{\collisiontext}{If $\CFshort = \Kershort$ then collision-free hash functions that can be evaluated in deterministic polynomial time do not exist.}

\newcommand{\collisionproof}{
The equivalence relation
$
\set{((i,x), (i,y)) : E(i,x) = E(i,y)}
$
has a canonical form $f \in \cc{FP}$ by hypothesis.  As in the proof of Proposition~\ref{prop:factoring}, the canonical form $f$ can be used by a randomized algorithm to find collisions in $h_{i}$ with non-negligible probability: choose $x$ at random, and if $f(x) \neq x$ then a collision has been found.

Since $h_{i}$ maps $\Sigma^{|i|+1} \to \Sigma^{|i|}$, there are at most $2^{|i|}-1$ singleton classes in $R=\Ker(h_{i})$.  If $x$ lies in an equivalence class of size at least $2$, then $\Pr_{x}[f(x) \neq x | \#[x]_{R} \geq 2] \geq \frac{1}{2}$.  Thus $\Pr_{x}[f(x) \neq x] = \Pr_{x}[f(x) \neq x | \#[x]_{R} \geq 2] \Pr_{x} [\#[x]_{R} \geq 2] \geq \frac{1}{2}\left(\frac{1}{2} + \frac{1}{2^{|i|+1}}\right) > \frac{1}{4}$.}

\newtheorem*{factoringprop}{Proposition~\ref{prop:factoring}}
\newcommand{\factoringtext}{If $\CFshort = \Kershort$ then integers can be factored in probabilistic polynomial time.}

\newcommand{\NPUPtext}{If $\CFshort = \Kershort$ then $\cc{NPMV_{g}} \subseteq_{c} \cc{NPSV_{g}}$.}

\newtheorem*{NPUPcor}{Corollary~\ref{cor:NPUP}}
\newcommand{\NPUPcortext}{If $\CFshort = \Kershort$ then $\cc{NP} = \cc{UP}$ and $\cc{PH} \subseteq \cc{S_{2}[NP \cap coNP]} \subseteq \cc{ZPP}\oracle{\cc{NP}}$.}

\newtheorem*{BQPthm}{Theorem~\ref{thm:BQP/RP}}
\newcommand{\BQPtext}{If $\Kershort = \Eqshort$ then $\cc{UP} \subseteq \cc{BQP}$.  If $\CFshort = \Eqshort$ then $\cc{UP} \subseteq \cc{RP}$.}

\newtheorem*{Promisethm}{Theorem~\ref{thm:promise}}
\newcommand{\Promisetext}{If $\cc{PromiseKer} = \cc{PromisePEq}$ then $\cc{NP} \subseteq \cc{BQP} \cap \cc{SZK}$, and in particular $\cc{PH} = \cc{AM}$.}

\newtheorem*{NPUPRPcor}{Corollary~\ref{cor:NPUPRP}}
\newcommand{\NPUPRPtext}{If $\CFshort = \Eqshort$ then $\cc{NP} = \cc{UP} = \cc{RP}$ and in particular, $\cc{PH}=\cc{BPP}$.}

\begin{document}

\title{Complexity Classes of Equivalence Problems Revisited}

\author{Lance Fortnow and Joshua A. Grochow}

\begin{abstract}
To determine if two lists of numbers are the same set, we sort both lists and see if we get the same result. The sorted list is a \emph{canonical form} for the equivalence relation of set equality. Other canonical forms arise in graph isomorphism algorithms. To determine if two graphs are cospectral (have the same eigenvalues), we compute their characteristic polynomials and see if they are equal; the characteristic polynomial is a \emph{complete invariant} for cospectrality. Finally, an equivalence relation may be decidable in $\cc{P}$ without either a complete invariant or canonical form.  Blass and Gurevich (SIAM J. Comput., 1984) ask whether these conditions on equivalence relations---having an $\cc{FP}$ canonical form, having an $\cc{FP}$ complete invariant, and being in $\cc{P}$---are distinct. They showed that this question requires non-relativizing techniques to resolve. We extend their results, and give new connections to probabilistic and quantum computation.
\end{abstract}

\maketitle


\section{Introduction}
Equivalence relations and their associated algorithmic problems arise throughout mathematics and computer science.  Examples run the gamut from trivial---decide whether two lists contain the same set of elements---to undecidable---decide whether two finitely presented groups are isomorphic \cite{novikov, booneVVI}.  Some examples are of great mathematical importance, and some are of great interest to complexity theorists, such as graph isomorphism ($\lang{GI}$).  

Complete invariants are a common tool for finding algorithmic solutions to equivalence problems.  Normal or canonical forms---where a unique representative is chosen from each equivalence class as the invariant of that class---are also quite common, particularly in algorithms for $\lang{GI}$ and its variants \cite{hopcroftWongPlanarIso, hopcroftTarjanPlanarIso, babaiLuks, furerSchnyderSpecker, millerBoundedGenusIso, babaiGrigoryevMount}.  More recently, Agrawal and Thierauf \cite{agrawalThierauf, thierauf} used a randomized canonical form to show that Boolean formula non-isomorphism ($\overline{\lang{FI}}$) is in $\cc{AM}\oracle{\cc{NP}}$.  More generally, the monograph by Thierauf \cite{thierauf} gives an excellent overview of equivalence and isomorphism problems in complexity theory.

Many efficient algorithms for special cases of $\lang{GI}$ have been upgraded to canonical forms or complete invariants.  Are these techniques necessary for an efficient algorithm?  Are these techniques distinct?  Gary Miller \cite{millerBoundedGenusIso} pointed out that $\lang{GI}$ has a polynomial-time complete invariant \ifftext it has a polynomial-time canonical form (see also \cite{gurevich}).  The general form of this question is central both in Blass and Gurevich \cite{blassGurevich1, blassGurevich2} and here: are canonical forms or complete invariants necessary for the efficient solution of equivalence problems?

In 1984, Blass and Gurevich \cite{blassGurevich1, blassGurevich2} introduced complexity classes to study these algorithmic approaches to equivalence problems.  Although we came to the same definitions and many of the same results independently, this work can be viewed partially as an update and a follow-up to their papers in light of the intervening 25 years of complexity theory.  The classes $\cc{UP}$, $\cc{RP}$, and $\cc{BQP}$, the function classes $\cc{NPMV}$ (multi-valued functions computed by $\cc{NP}$ machines) and $\cc{NPSV}$ (single-valued functions computed by $\cc{NP}$ machines), and generic oracle (forcing) methods feature prominently in this work.

Blass and Gurevich \cite{blassGurevich1, blassGurevich2} introduced the following four problems and the associated complexity classes.  Where they use ``normal form'' we say ``canonical form,'' though the terms are synonymous and the choice is immaterial.  We also introduce new notation for these complexity classes that makes the distinction between language classes and function classes more explicit.  For an equivalence relation $R \subseteq \Sigma^{*} \times \Sigma^{*}$, they defined:

The \definedWord{recognition problem}: given $x,y \in \Sigma^{*}$, decide whether $x \sim_{R} y$.

The \definedWord{invariant problem}: for $x \in \Sigma^{*}$, calculate a complete invariant $f(x) \in \Sigma^{*}$ for $R$, that is, a function such that $x \sim_{R} y$ \ifftext $f(x) = f(y)$.

The \definedWord{canonical form problem}: for $x \in \Sigma^{*}$ calculate a canonical form $f(x) \in \Sigma^{*}$ for $R$, that is, a function such that $x \sim_{R} f(x)$ for all $x \in \Sigma^{*}$, and $x \sim_{R} y$ implies $f(x) = f(y)$.

The \definedWord{first canonical form problem}: for $x \in \Sigma^{*}$, calculate the first $y \in \Sigma^{*}$ such that $y \sim_{R} x$.  Here, ``first'' refers to the standard length-lexicographic ordering on $\Sigma^{*}$, though any ordering that can be computed easily enough would suffice.

The corresponding polynomial-time complexity classes are defined as follows:

\begin{defn}
$\Eqshort$ consists of those equivalence relations whose recognition problem has a polynomial-time solution.  $\ccKer{FP}$ consists of those equivalence relations that have a polynomial-time computable complete invariant.  $\ccCF{FP}$ consists of those equivalence relations that have a polynomial-time canonical form.  $\ccLexEq{FP}$ consists of those equivalence relations whose first canonical form is computable in polynomial time.
\end{defn}

We occasionally omit the ``$\cc{FP}$'' from the latter three classes.  It is obvious that 
\[
\FCFshort \subseteq \CFshort \subseteq \Kershort \subseteq \Eqshort,
\]
and our first guiding question is: which of these inclusions is tight?

\subsection{Examples}
To get a better feel for these complexity classes and help motivate them, we begin with several examples, especially including those that potentially witness the separation of these classes. Some of these will be discussed in more depth in Section~\ref{sec:hardProblems}. We also rephrase some of the examples we have already mentioned using these classes.

\begin{example} Graph isomorphism is in $\cc{NPEq}$ (equivalence problems decidable in $\cc{NP}$), and is in $\ccKer{FP}$ if and only if it is in $\ccCF{FP}$ \cite{millerBoundedGenusIso} (see also \cite{gurevich}). In fact, this result also holds for any function class that is closed under $\cc{FP}$ reductions such as $\cc{FP}\oracle{\cc{NP} \cap \cc{coNP}}$. \end{example}

\begin{example} \label{ex:formulaEquiv}
Boolean formula equivalence (do two Boolean formulae compute the same function) is in $\cc{coNPEq}$, and is $\cc{coNP}$-complete (to check if $\varphi$ is a tautology, see if it is equivalent to the constant-true formula $1$).
\end{example}

\begin{example} Sorting a list is a first canonical form for set equality. Set equality is thus in $\ccLexEq{FP}$. \end{example}

\begin{example} The characteristic polynomial is a polynomial-time complete invariant for graph cospectrality. No polynomial-time canonical form is known for this problem, so graph cospectrality is a potential witness to $\CFshort \neq \Kershort$. \end{example}

\begin{example} \label{ex:subgroupEquiv}
The \definedWord{subgroup equality problem} is: given two subsets $\set{g_{1},\dotsc,g_{t}}$, $\set{h_{1},\dotsc,h_{s}}$ of a group $G$ determine if they generate the same subgroup. For permutation groups on $\set{1, \dotsc, n}$, this problem lies in $\ccCF{FP}$, via a simple modification \cite{babaiPersonalComm} of the classic techniques of Sims \cite{sims1970, sims1971}, whose analysis was completed by Furst, Hopcroft, and Luks \cite{furstHopcroftLuks} and Knuth \cite{knuth}. However, the subgroup equality problem for other groups is a potential source of witnesses to $\Kershort \neq \Eqshort$. 
\end{example}

Although factoring integers is not an equivalence problem, its hardness would imply $\CFshort \neq \Kershort$, as the next proposition shows. In Section~\ref{sec:collision-free}, we show a similar result based on the hardness of collision-free hash functions that can be computed deterministically. The proof of this proposition highlights what seems to be an essential difference between $\CFshort$ and $\Kershort$.

\begin{prop}\orig
\label{prop:factoring}
\factoringtext 
\end{prop}

\begin{proof}
Suppose we wish to factor an integer $N$.  We may assume $N$ is not prime, since primality can be determined in polynomial time \cite{aksPrimality}, but even much weaker machinery lets us do so in probabilistic polynomial time \cite{solovayStrassen,rabin:primality}, which is sufficient here.  By hypothesis, the kernel of the Rabin function $x \mapsto x^{2} \pmod{N}$:
\[
R_{N} = \set{(x,y) : x^{2} \equiv y^{2} \pmod{N}}
\]
has a canonical form $f \in \cc{FP}$.

Randomly choose $x \in \Z/N\Z$ and let $y=f(x)$.  Then $x^{2} \equiv y^{2} \pmod{N}$; equivalently, $(x-y)(x+y) \equiv 0 \pmod{N}$.  If $y \not\equiv \pm x \pmod{N}$, then since neither $x-y$ nor $x+y$ is $\equiv 0\pmod{N}$, $\gcd(N, x-y)$ is a nontrivial factor $z$ of $N$.  Let $r(N)$ be the least number of distinct square roots modulo $N$.  Then $\Pr_x[y \not\equiv \pm x] \geq 1-\frac{2}{r(N)}$.  Since $N$ is composite and odd without loss of generality, $r(N) \geq 4$.  Thus $\Pr_x[y \not\equiv \pm x] = \Pr_x[\text{the algorithm finds a factor of $N$}] \geq \frac{1}{2}$.  Recursively call the algorithm on $N/z$.
\end{proof}

\subsection{Main results}

Blass and Gurevich showed that none of the four problems above polynomial-time Turing-reduces (Cook-reduces) to the next in line.  We extend their results using generic oracles, and we also give further complexity-theoretic evidence for the separation of these classes, giving new connections to probabilistic and quantum computing.  Our main results in this regard are:

\begin{factoringprop}
\textit{\factoringtext}
\end{factoringprop}

\begin{collisionprop}\orig
\textit{\collisiontext}
\end{collisionprop}

\begin{BQPthm}\orig
\textit{\BQPtext}
\end{BQPthm}

\begin{Promisethm}\orig
\textit{\Promisetext}
\end{Promisethm}

We give the definitions of $\cc{PromisePEq}$ and $\cc{PromiseKer}$ in Section~\ref{sec:promise}.  We also show the following two related results:

\begin{NPUPcor}
\textit{\NPUPcortext}
\end{NPUPcor}

\begin{NPUPRPcor}\orig
\textit{\NPUPRPtext}
\end{NPUPRPcor}

\noindent Corollary~\ref{cor:NPUP} follows from the slightly stronger Theorem~\ref{thm:NPUP}, but we do not give the statement here as it requires further definitions.


\subsection{Organization}

The remainder of the paper is organized as follows.  In Section \ref{sec:prelims} we give preliminary definitions and background.  In Section \ref{sec:previousResults} we review the original results of Blass and Gurevich \cite{blassGurevich1, blassGurevich2}.  We also combine their results with other results that have appeared in the past 25 years to yield some immediate extensions.
In Section \ref{sec:probQuantum} we prove new results connecting these classes with probabilistic and quantum computation.  
In Section \ref{sec:promise} we introduce the promise versions of $\Eqshort$ and $\Kershort$ and prove Theorem~\ref{thm:promise}.
In Section \ref{sec:groupy}, we introduce a group-like condition on the witness sets of $\cc{NP}$-complete problems that would allow us to extend the first half of Theorem~\ref{thm:BQP/RP} from $\cc{UP}$ to $\cc{NP}$, giving much stronger evidence that $\Kershort \neq \Eqshort$.  We believe the question of whether any $\cc{NP}$-complete sets have this property is of independent interest: a positive answer would provide nontrivial quantum algorithms for $\cc{NP}$ problems, and a negative answer would provide further concrete evidence for the lack of structure in $\cc{NP}$-complete problems.  
In Section \ref{sec:hardProblems} we discuss collision-free hash functions, the subgroup equality problem and Boolean function congruence (not isomorphism) as potential witnesses to the separation of these classes. We also introduce a notion of reduction between equivalence relations and the corresponding notion of completeness.
In Section \ref{sec:oracles}, we update and extend some of the oracle results of Blass and Gurevich \cite{blassGurevich1, blassGurevich2} using generic oracles.
In the final section we mention several directions for further research, in addition to the several open questions scattered throughout the paper.
  
\section{Preliminaries} \label{sec:prelims}
We assume the reader is familiar with standard complexity classes such as $\cc{P}$, $\cc{NP}$, $\cc{BPP}$, and the polynomial hierarchy $\cc{PH} = \bigcup \cc{\Sigma_{k} P} = \bigcup \cc{\Pi_{k} P} = \bigcup \cc{\Delta_{k} P}$.  We refer the reader to the textbook by Arora and Barak \cite{aroraBarak} and the Complexity Zoo at \url{http://qwiki.stanford.edu/index.php/Complexity_Zoo} for more details.

A language $L$ is in the class $\cc{UP}$ if there is a nondeterministic machine deciding $L$ that has at most one accepting path on each input.

The class $\cc{BQP}$ consists of those languages that can be decided on a quantum computer in polynomial time with error strictly bounded away from $1/2$.  For more details on quantum computing, we recommend the book by Nielson and Chuang \cite{nielsonChuang}.

For any class $\mathcal{C}$, the class $\cc{S_{2}[\mathcal{C}]}$ is defined as follows.  A language $L$ is in $\cc{S_{2}[\mathcal{C}]}$ if there is a language $V \in \mathcal{C}$ and a polynomial $p$ such that
\begin{eqnarray*}
x \in L & \implies & (\exists y : |y| \leq p(|x|))(\forall z : |z| \leq p(|x|))[V(x,y,z) = 1] \\
x \notin L & \implies & (\exists z : |z| \leq p(|x|))(\forall y : |y| \leq p(|x|))[V(x,y,z) = 0].
\end{eqnarray*}
The class $\cc{S_{2} P}$ was defined independently by Russell and Sundaram \cite{russellSundaramBPP} and Canetti \cite{canettiBPP}.  Cai \cite{caiS2P} showed that $\cc{S_{2}[NP \cap coNP]} \subseteq \cc{ZPP}\oracle{\cc{NP}}$.

\subsection{Function Classes}
Complexity-bounded function classes are defined in terms of Turing transducers.  A transducer only outputs a value if it enters an accepting state.  In general, then, a nondeterministic transducer can be partial and/or multi-valued.  For such a function $f$, we write
\[
set\dash f(x) = \set{y : \text{ some accepting computation of $f(x)$ outputs $y$}}
\]
The \definedWord{domain} of a partial multi-valued function is the set
\[
\dom(f) = \set{x : set\dash f(x) \neq \emptyset}.
\]
The \definedWord{graph} of a partial multi-valued function is the set
\[
\grph(f) = \set{(x, y) : y \in set\dash f(x)}.
\]

The class $\cc{FP}$ is the class of all total functions computable in deterministic polynomial time.  The class $\cc{\partialFP}$ is the class of all partial functions computable in deterministic polynomial time.  Note that machines computing a $\cc{\partialFP}$ function must halt in polynomial time even when they make no output.

The class $\cc{FL}$ is the class of all total functions computable by deterministic logarithmic-space transducers, that is, the length of the output and the $i$-th bit of the output of the function can be computed in logarithmic-space.

The class $\cc{NPSV}$ consists of all single-valued partial functions computable by a nondeterministic polynomial-time transducer.  Note that multiple branches of an $\cc{NPSV}$ transducer may accept, but they must all have the same output.
The class $\cc{NPMV}$ consists of all multi-valued partial functions computable by a nondeterministic polynomial-time transducer.
The classes $\cc{NPSV_{t}}$ and $\cc{NPMV_{t}}$ are the subclasses of $\cc{NPSV}$ and $\cc{NPMV}$, respectively, consisting of the total functions in those classes.
The classes $\cc{NPSV_{g}}$ and $\cc{NPMV_{g}}$ are the subclasses of $\cc{NPSV}$ and $\cc{NPMV}$, respectively, whose graphs are in $\cc{P}$.

A \definedWord{refinement} of a multi-valued partial function $f$ is a multi-valued partial function $g$ such that $\dom(g) = \dom(f)$ and $set\dash g(x) \subseteq set\dash f(x)$ for all $x$.  In particular,  if $set\dash f(x)$ is nonempty then so is $set\dash g(x)$.
If $\mathcal{F}_1$ and $\mathcal{F}_{2}$ are two classes of partial multi-valued functions, then
\[
\mathcal{F}_1 \subseteq_{c} \mathcal{F}_{2}
\]
means that every function in $\mathcal{F}_1$ has a refinement in $\mathcal{F}_{2}$.

It is known that $\cc{NPMV} \subseteq_{c} \cc{\partialFP}$ \ifftext $\cc{P} = \cc{NP}$ \cite{selman1992} \ifftext $\cc{NPSV} \subseteq \cc{\partialFP}$ \cite{sxb83}.  Selman \cite{selmanFunctions} is one of the classic works in this area, and gives many more results regarding these function classes.

\subsection{Equivalence Relations} \label{sec:equiv}
For an equivalence relation $R \subseteq \Sigma^{*} \times \Sigma^{*}$, we write $x \sim_{R} y$ if $(x,y) \in R$.  We write $[x]_{R}$ for the $R$-equivalence class of $x$.  The \definedWord{kernel} of a function $f$ is the equivalence relation $\Ker(f) = \set{(x,y) : f(x)=f(y)}$.  For an equivalence relation $R$, if $R = \Ker(f)$, we say that $f$ is a \definedWord{complete invariant} for $R$.  If, furthermore, $x \sim_{R} f(x)$ for every $x$, then $f$ is a \definedWord{canonical form} for $R$.  If, further still, $f(x)$ is the first member of $[x]_{R}$ under lexicographic order, we say that $f$ is the \definedWord{first canonical form} for $R$.  The \definedWord{trivial relation} is all of $\Sigma^{*} \times \Sigma^{*}$, that is, all strings are equivalent under the trivial relation, or equivalently $[x]=\Sigma^{*}$ for all $x$.  

An equivalence relation is \definedWord{length-restricted} if $x \sim y$ implies $|x| = |y|$.  An equivalence relation is \definedWord{polynomially bounded} if there is a polynomial $p$ such that $x \sim y$ implies $|x| \leq p(|y|)$.  Note that the first canonical form for a polynomially bounded equivalence relation is a polynomially honest function.  If $\mathcal{C}$ is a class of equivalence relations, we write $\mathcal{C}_{=}$ for the class of length-restricted equivalence relations in $\mathcal{C}$, and $\mathcal{C}_{p}$ for the class of polynomially bounded equivalence relations in $\mathcal{C}$.

Let $\tuple{\cdot, \cdot} \colon \Sigma^{*} \times \Sigma^{*} \to \Sigma^{*}$ be a polynomial-time computable and polynomial-time invertible pairing function such that $|\tuple{x, y}|$ depends only on $|x|$ and $|y|$. By polynomial-time invertible we mean that the projection functions $\pi_{i}(\tuple{x_{1}, x_{2}}) = x_{i}$ for $i=1,2$ are computable in polynomial time.

\section{Previous Results} \label{sec:previousResults}
Here we recall the previous results most relevant to our work.  Most of the results in this section are from Blass and Gurevich \cite{blassGurevich1, blassGurevich2}. We are not aware of any other prior work in this area.  However, results in other areas of computational complexity that have been obtained since 1984 can be used as black boxes to extend their results, which we do here.

We mention that analogues of these classes for finite-state machines have been studied, and nearly all their interrelationships completely determined \cite{automataEq}. For the class of computable functions or the class of primitive recursive functions, Blass and Gurevich \cite{blassGurevich1} already noted that all four classes of equivalence relations are equal.

If $R \in \cc{PEq}$, then the language $R' = \set{(x,y) : (\exists z)[z \leq_{lex} y \text{ and } (x,z) \in R]}$ is in \cc{NP}, and can be used to perform a binary search for the first canonical form for $R$.  Hence, $\cc{PEq} \subseteq \ccLexEq{FP\oracle{\cc{NP}}}$.  The first result shows that this containment is tight:

\begin{thm}[\cite{blassGurevich1} Theorem~1] There is an equivalence relation $R \in \CFshort$ whose first canonical form problem is essentially $\cc{\Delta_{2} P}$-complete, that is, it is in $\cc{FP\oracle{\cc{NP}}}=\cc{F \Delta_{2} P}$ and is $\cc{\Delta_{2} P}$-hard. \end{thm}

Note that the above proof that $\cc{PEq} \subseteq \ccLexEq{FP\oracle{\cc{NP}}}$ relativizes, so all four polynomial-time classes of equivalence relations are equal in any world where $\cc{P} = \cc{NP}$, in particular, relative to any $\cc{PSPACE}$-complete oracle.  The next result gives relativized worlds in which $\Kershort \neq \Eqshort$, $\CFshort \neq \Kershort$, and $\FCFshort \neq \CFshort$, though these worlds cannot obviously be combined.

\begin{thm}[Blass \& Gurevich \cite{blassGurevich1} Theorem~2] \label{thm:BGreducible} Of the four equivalence problems defined above, none is Cook reducible to the next in line.  In particular:

\begin{enumerate}
\renewcommand{\theenumi}{\alph{enumi}}
\item\label{thm:BGreducible-invrec}  There is an equivalence relation $R \notin \ccKer{FP\oracle{R}}$, \ie, $\ccKer{FP\oracle{R}} \neq \cc{P\oracle{R} Eq}$.

\item There is a function $f$ such that $\Ker(f) \notin \ccCF{FP\oracle{f}}$, \ie, $\ccCF{FP\oracle{f}} \neq \ccKer{FP\oracle{f}}$.

\item There is an idempotent function $f$ such that $\Ker(f) \notin \ccLexEq{FP\oracle{f}}$, \ie, $\ccLexEq{FP\oracle{f}} \neq \ccCF{FP\oracle{f}}$.

\end{enumerate}
Furthermore, there is an equivalence relation $R \notin \ccKer{NPSV_{t}\oracle{R}}$, \ie, $\cc{P\oracle{R}Eq} \not\subseteq \ccKer{NPSV_{t}\oracle{R}}$ \cite[Thm.~5]{blassGurevich2}. \end{thm}

In addition to several extensions of these results, Blass and Gurevich \cite{blassGurevich1, blassGurevich2} also show that collapses between certain classes of equivalence problems are equivalent to more standard complexity-theoretic hypotheses.  Here we collect some of their main results:

\begin{thm} \label{thm:someBGResults}
\begin{enumerate}
\renewcommand{\theenumi}{\arabic{enumi}}
\item $\ccCF{FP} \subseteq \ccLexEq{NPSV_{t}} \Iff \cc{NPEq} \subseteq \cc{coNPEq} \Iff \cc{coNPEq} \subseteq \cc{NPEq} \Iff \cc{NP} = \cc{coNP}$ \cite[Thm.~1]{blassGurevich2}.

\item $\ccLexEq{NPSV_{t}} \subseteq \cc{PEq} \Iff \cc{P} = \cc{NP} \cap \cc{coNP}$ \cite[Thm.~2]{blassGurevich2}.

\end{enumerate}
\end{thm}

Note that $\cc{NPEq}$ consists of those equivalence relations decidable in $\cc{NP}$, and is distinct from $\cc{P\oracle{\cc{NP}}Eq}$ assuming $\cc{NP} \neq \cc{P}\oracle{\cc{NP}}$. This follows from the observation that, for any set $A$ there is an equivalence relation $R$ that is polynomial-time equivalent to $A$, namely the equivalence relation generated by $\set{(0x, 1x) : x \in A}$ (if $A$ is neither empty nor $\Sigma^{*}$, then $A \equiv_{m}^{p} R$; in any case,  $A \equiv_{1-tt}^{p} R$). 

We think the following result is one of their most surprising:

\begin{thm}[Blass \& Gurevich \cite{blassGurevich2} Theorem~3] \label{thm:BGuniform} The following statements are equivalent:
\begin{enumerate}
\renewcommand{\theenumi}{\arabic{enumi}}
\item $\ccKer{FP}_{=} \subseteq \ccCF{NPSV_{t}}$.
\item $\cc{NP}$ has the shrinking property (see Gla{\ss}er, Reitwie{\ss}ner, and Selivanov \cite{shrinking}): if $A, B \in \cc{NP}$, then there are disjoint $A', B' \in \cc{NP}$ such that $A' \subseteq A$, $B' \subseteq B$, and $A \union B = A' \union B'$.
\item $\cc{NPMV} \subseteq_{c} \cc{NPSV}$, \ie, the uniformization principle holds for \cc{NP}.
\end{enumerate}
\end{thm}

Hemaspaandra, Naik, Ogihara, and Selman \cite{HNOS96} showed that if $\cc{NPMV} \subseteq_{c} \cc{NPSV}$ then $\cc{SAT} \in (\cc{NP} \cap \cc{coNP})/poly$.  At the time, the strongest known consequence of $\cc{SAT} \in (\cc{NP} \cap \cc{coNP})/poly$ was $\cc{PH} = \cc{\Sigma_{2} P}$ \cite{karpLipton}.  Shortly thereafter K\"{o}bler and Watanabe \cite{koblerWatanabeFull} improved the collapse to $\cc{PH} = \cc{ZPP}\oracle{\cc{NP}}$, and in the early 2000's Cai, Chakaravarthy, Hemaspaandra, and Ogihara \cite{CCHO} further improved the collapse to $\cc{PH} = \cc{S_{2}[NP \cap coNP]}$.  Combined with Theorem~\ref{thm:BGuniform}, this immediately implies a result that has not been announced previously:

\begin{cor} \label{cor:PH=ZPPNP}
If $\CFshort = \Kershort$ then $\cc{PH} \subseteq \cc{S_{2}[NP \cap coNP]} \subseteq \cc{ZPP}\oracle{\cc{NP}}$. \qed
\end{cor}

\section{Evidence for Separation}

\subsection{New Collapses} \label{sec:probQuantum}

Blass and Gurevich's \cite{blassGurevich2} proof that $\ccKer{FP}_{=} \subseteq \ccCF{NPSV_{t}} \implies \cc{NPMV} \subseteq_{c} \cc{NPSV}$ essentially shows the following slightly stronger result.  However, as $\cc{NPMV} \subseteq_{c} \cc{NPSV}$ is not known to imply $\cc{NPMV_{g}} \subseteq_{c} \cc{NPSV_{g}}$, our result does not directly follow from their \emph{result}, but only from its proof, the core of which is reproduced here:

\begin{thm}
\label{thm:NPUP}
\NPUPtext
\end{thm}

\begin{proof}
Let $f \in \cc{NPMV_{g}}$, let $M$ be a nondeterministic polynomial-time transducer computing $f$, and let $V$ be a polynomial-time decider for $\grph(f)$.  If $\CFshort = \Kershort$, then the equivalence relation
\[
\set{((x,y), (x,y')) : V(x,y) = V(x,y')} = \Ker( (x,y) \mapsto (x, V(x,y)))
\]
has a canonical form $c \in \cc{FP}$.  Then the following algorithm computes a refinement of $f$ in $\cc{NPSV_{g}}$: simulate $M(x)$.  On each branch, if the output would be $y$, accept \ifftext $c(x,y) = (x,y)$.  Hence $f \in_{c} \cc{NPSV_{g}}$.
\end{proof}

Similar to the original result \cite{blassGurevich2}, we can weaken the assumption of this theorem to \newline $\Kershort_{p} \subseteq \CFshort$, without modifying the proof.  By padding, we can further weaken the assumption to $\Kershort_{=} \subseteq \CFshort$. 

\begin{cor}
\label{cor:NPUP}
\NPUPcortext \qed
\end{cor}

Note that Corollary~\ref{cor:PH=ZPPNP} alone does not imply Corollary~\ref{cor:NPUP}, as neither of the statements $\cc{PH} = \cc{S_{2}[NP \cap coNP]}$ and $\cc{NP} = \cc{UP}$ is known to imply the other.  Indeed, it is still an open question as to whether $\cc{NP} = \cc{UP}$ implies any collapse of $\cc{PH}$ whatsoever. 

The next new result we present gives a new connection between complexity classes of equivalence problems and quantum and probabilistic computation:

\begin{thm}\orig
\label{thm:BQP/RP}
\BQPtext
\end{thm}

\begin{proof}
Suppose $\Kershort = \Eqshort$.  Let $L$ be a language in \cc{UP}, let $V$ be a $\cc{UP}$ verifier for $L$, let $p$ be a polynomial bounding the size of $V$-witnesses for $L$.  Consider the relation
\[
R_{L} = \set{((a,x),(a,y)) : x=y \text{ or } |x| = |y| \mbox{ and } V(a,x \oplus y)=1 }
\]
where $\oplus$ denotes bit-wise exclusive-or.  Clearly $R_{L} \in \Eqshort$, so by hypothesis $R_{L}$ has a complete invariant $f \in \cc{FP}$.  Since $L \in \cc{UP}$, for each $a \in L$ there is a unique string $w_{a}$ such that $V(a,w_{a})=1$.  Define $f_{a}(x) = f(a,x)$.  Then for all distinct $x$ and $x'$, $f_{a}(x) = f_{a}(x')$ \ifftext $x \oplus x' = w_{a}$.  Given $a$ and $f_{a}$, and the promise that $f_{a}$ is either injective or two-to-one in the manner described, finding $w_{a}$ or determining that there is no such string is exactly Daniel Simon's problem, which is in \cc{BQP} \cite{simon97}.

Now suppose further that $\CFshort = \Eqshort$.  Then we may take $f$ to be not only a complete invariant but further a canonical form for $R_{L}$.  On input $a$, the following algorithm decides $L$ in polynomial time with bounded error: for each length $\ell \leq p(|a|)$, pick a string $x$ of length $\ell$ at random, compute $f((a,x))=(a,y)$, and compute $V(a,x \oplus y)$.  If $V(a,x \oplus y)=1$ for any length $\ell$, output 1.  Otherwise, output 0.  If $a \notin L$ then this algorithm always returns 0.  If $a \in L$ and $0^{\ell}$ is $a$'s witness, then the algorithm always returns $1$.  If $a \in L$ and $0^{\ell}$ is not $a$'s witness, then $y \neq x$, and hence the answer is correct, with probability $1/2$. \end{proof}

We would like to extend the first half of Theorem~\ref{thm:BQP/RP} from \cc{UP} to \cc{NP} to give stronger evidence that $\Kershort \neq \Eqshort$, but the techniques do not obviously apply.  We pose two approaches to this problem in Sections \ref{sec:promise} and \ref{sec:groupy}.

\begin{cor}\orig
\label{cor:NPUPRP}
\NPUPRPtext
\end{cor}

\begin{proof}
If $\CFshort = \Eqshort$ then it follows directly from Theorems \ref{thm:NPUP} and \ref{thm:BQP/RP} that $\cc{NP} = \cc{UP} \subseteq \cc{RP}$.  Thus $\cc{NP} = \cc{RP}$, since $\cc{RP} \subseteq \cc{NP}$ without any assumptions.  Furthermore, it follows that $\cc{PH} \subseteq \cc{BPP}$ \cite{zachosPHBPP}, and since $\cc{BPP} \subseteq \cc{PH}$ \cite{lautemannBPP, sipserBPP}, the two are equal.
\end{proof}

The collapse inferred here is stronger than that of Corollary~\ref{cor:PH=ZPPNP}, since $\cc{BPP} \subseteq \cc{S_{2} P} \subseteq \cc{S_{2}[NP \cap coNP]}$ \cite{russellSundaramBPP, canettiBPP}.  However, this result is incomparable to Corollary~\ref{cor:PH=ZPPNP} since it also makes the stronger assumption $\CFshort = \cc{PEq}$, rather than only assuming $\CFshort = \Kershort$.

\subsubsection{Promise classes} \label{sec:promise}

One way to extend the first half of Theorem~\ref{thm:BQP/RP} from \cc{UP} to \cc{NP}, suggested to us by Scott Aaronson \cite{aaronsonPersonalComm}, involves promise versions of $\Eqshort$ and $\Kershort$.  

\begin{defn}
A language $R$ of triples is in $\cc{PromisePEq}$ if there is a polynomial-time algorithm $A$ such that, whenever $R_{a} = \set{(x, y) : (a, x, y) \in R}$ is an equivalence relation, $A(a,x,y) = R(a,x,y)$ for all $x, y \in \Sigma^{*}$.

Similarly, $R$ is in $\cc{PromiseKer}$ if there is a polynomial-time function $f$ such that, whenever $R_{a}$ is an equivalence relation, $f(a,x) = f(a,y) \iff (a,x,y) \in R$ for all $x, y \in \Sigma^{*}$.  We call such $f$ a \definedWord{promise complete invariant} for $R$.
\end{defn}

As usual for promise classes, if $R_{a}$ is not an equivalence relation, we do not restrict the
output of $A(a,x,y)$ or $f(a,x)$ in any way.

\begin{thm}\orig
\label{thm:promise}
\Promisetext
\end{thm}

\begin{proof}
The first part of the proof follows that of Theorem~\ref{thm:BQP/RP}, treating the promises with care.  Suppose $\cc{PromiseKer} = \cc{PromisePEq}$.  Let $L$ be a language in $\cc{PromiseUP}$, let $V$ be a $\cc{PromiseUP}$ verifier for $L$, let $p$ be a polynomial bounding the size of $V$-witnesses for $L$.  That is, if $\# V(x) = \# \set{y : V(x,y) = 1} \leq 1$ then $x \in L \iff (\exists y)[|y| \leq p(|x|) \text{ and } V(x,y)=1]$.  Consider the relation
\[
R_{L} = \set{((a,x),(a,y)) : x=y \text{ or } |x| = |y| \mbox{ and } V(a,x \oplus y)=1 }
\]
(the same relation as in Theorem~\ref{thm:BQP/RP}).  Clearly $R_{L} \in \cc{PromisePEq}$, so by hypothesis $R_{L}$ has a promise complete invariant $f \in \cc{FP}$.  Since $L \in \cc{PromiseUP}$, for each $a \in L$ such that $\# V(x) = 1$, there is a unique string $w_{a}$ such that $V(a,w_{a})=1$.  Define $f_{a}(x) = f(a,x)$.  Then for all distinct $x$ and $x'$, $f_{a}(x) = f_{a}(x')$ \ifftext $x \oplus x' = w_{a}$.  As in Theorem~\ref{thm:BQP/RP}, given $a$ and $f_{a}$, finding $w_{a}$ or determining that there is no such string is exactly Simon's problem, which is in \cc{BQP} \cite{simon97}.  Here, of course, we have reduced to the \emph{promise version} of Simon's problem.

To show $\cc{NP} \subseteq \cc{BQP}$, we use the technique of Valiant and Vazirani \cite{valiantVazirani}: given a Boolean formula $\varphi$, they randomly produce a formula $\varphi'$ such that if $\varphi$ is unsatisfiable, then so is $\varphi'$, and if $\varphi$ is satisfiable, then $\varphi'$ has a \emph{unique} satisfying assignment with probability at least $1/p(|\varphi|)$ for some polynomial $p$.  In this case, $(\varphi', f_{\varphi'})$ satisfies the promise of Simon's problem, and the $\cc{BQP}$ algorithm for Simon's problem either finds the satisfying assignment to $\varphi'$ or correctly reports that none exists.  Since the initial randomized construction of $\varphi'$ from $\varphi$ can also be carried out in $\cc{BQP}$, this whole algorithm puts $\lang{SAT} \in \cc{BQP}$.

Next we show $\cc{NP} \subseteq \cc{SZK}$.  As above, we randomly transform a Boolean formula $\varphi$ into a formula $\varphi'$ which has at most one satisfying assignment, with probability at least $1/p(|\varphi|)$.  Then we run the $\cc{SZK}$ protocol for Simon's problem on $\varphi'$, which we reproduce here for completeness.  If $\varphi'(00\dotsb 0) = 1$, then the verifier accepts immediately. Otherwise, the verifier randomly picks $x$ and sends $f_{\varphi'}(x) = f(\varphi', x)$ to the prover; the prover must try to recover $x$.  If $\varphi'$ has no satisfying assignments, then $f_{\varphi'}$ is one-to-one, and the prover always succeeds.  If $\varphi'$ has a (unique, not-all-zero) satisfying assignment, then $f_{\varphi'}$ is two-to-one, and the prover fails with probability at least $1/2$.  It is clear that this is an $\cc{SZK}$ protocol.

Since the construction of $\varphi'$ from $\varphi$ does not require any interaction between the prover and verifier, it can be prepended to the above protocol to give a statistical zero-knowledge protocol for $\lang{SAT}$.

Finally, we have $\cc{SZK} \subseteq \cc{AM} \cap \cc{coAM}$ \cite{fortnowSZK, aielloHastad}, and $\cc{NP} \subseteq \cc{coAM}$ implies $\cc{PH} = \cc{AM}$ \cite{babaiAM, boppanaHastadZachos}.
\end{proof}

The two conclusions of the above theorem (that is, ``$\cc{NP} \subseteq \cc{BQP}$'' and ``$\cc{PH} = \cc{AM}$'') are not known to be related by implication in either direction.  Even $\cc{NP} \subseteq \cc{BQP}$ and $\cc{NP} \subseteq \cc{SZK}$ are not known to be related by implication.  Indeed, there is an oracle relative to which $\cc{SZK}$ is not contained in $\cc{BQP}$ \cite{aaronsonSZKBQP}, and there is an oracle relative to which $\cc{BQP}$ is not contained in $\cc{SZK}$ \cite{CCDFGS}.

\subsubsection{Groupy witnesses for \cc{NP} problems} \label{sec:groupy}

The technique of the first half of Theorem~\ref{thm:BQP/RP} does not apply to arbitrary problems in $\cc{NP}$.  However, if an $\cc{NP}$ problem's witnesses satisfy a certain group-like condition, then Theorem~\ref{thm:BQP/RP} may be extended to that problem.  

Let $L \in \cc{NP}$ and let $V$ be a polynomial-time verifier for $L$.  By padding if necessary, we may suppose that for each $a \in L$, $a$'s witnesses all have the same length.  Suppose there is a polynomial-time length-restricted group structure on $\Sigma^{*}$, that is, a function $f \in \cc{FP}$ such that for each length $n$, $\Sigma^{n}$ is given a group structure defined by $xy^{-1} \defeq f(x,y)$.  Then
\[
R_{L} = \set{((a,x), (a,y)) : x=y \text{ or } V(a,xy^{-1})=1}
\]
is an equivalence relation \ifftext $a$'s witnesses are a subgroup of this group structure, or a subgroup less the identity.  The technique of Theorem~\ref{thm:BQP/RP} then reduces $L$ to the hidden subgroup problem over the family of groups defined by $f$.

The \emph{hidden subgroup problem}, or HSP, for a group $G$ is: given generators for $G$, an oracle computing the operation $(x,y) \mapsto xy^{-1}$, a set $X$, and a function $f\colon G \to X$ such that $\Ker(f)$ is the partition given by the right cosets of some subgroup $H \leq G$, find a generating set for $H$ \cite{kitaev}.  Hidden subgroup problems have played a central role in the study of quantum algorithms.  Integer factoring and the discrete logarithm problem both easily reduce to abelian HSPs.  The first polynomial-time quantum algorithm for these problems was discovered by Shor \cite{shor}; Kitaev \cite{kitaev} then noticed that Shor's algorithm in fact solves all abelian HSPs.  The unique shortest vector problem for lattices reduces to the dihedral HSP \cite{regevDHSPFinal}, which is solvable in subexponential quantum time \cite{kuperbergDHSP}.  The graph isomorphism problem reduces to the HSP for the symmetric group \cite{bealsQuantum} or the wreath product $S_{n} \wr S_{2}$ \cite{ettingerHoyer}, but it is still unknown whether any nontrivial quantum algorithm exists for $\lang{GI}$.  


The proof of Theorem~\ref{thm:BQP/RP} showed that if $\Kershort = \Eqshort$ then every language in $\cc{UP}$ reduces to Daniel Simon's problem.  We can now see that Simon's problem is in fact the HSP for $(\Z/2\Z)^{n}$, where the hidden subgroup has order $2$.  Simon \cite{simon97} gave a zero-error expected polynomial-time quantum algorithm for this problem, putting it in $\cc{ZQP} \subseteq \cc{BQP}$.  This result was later improved by Brassard and H{\o}yer \cite{brassardHoyer97} to a worst-case polynomial time quantum algorithm, that is, in the class $\cc{EQP}$ (sometimes referred to as just $\cc{QP}$).

This discussion motivates the following definition, results, and open question:

\begin{defn}\orig Let $L \in \cc{NP}$.  For each $a$ let $W(a)$ denote the set of $a$'s witnesses; without loss of generality, by padding if necessary, assume that $W(a) \subseteq \Sigma^{n}$ for some $n$.  The language $L$ has \definedWord{groupy witnesses} if there are functions $\text{mul},\text{gen},\text{dec} \in \cc{FP}$ such that for each $a \in L$:
\begin{enumerate}
\item let $G(a) = \set{x \in \Sigma^{n} : \text{dec}(a,x) = 1}$; then for all $x,y \in G(a)$, defining $xy^{-1} \defeq \text{mul}(a,x,y)$ gives a group structure to $G(a)$;
\item $\text{gen}(a)=(g_{1}, g_{2}, \dotsc, g_{k})$ is a generating set for $G(a)$; and 
\item $W(a)$ is a subgroup of $G(a)$, or a subgroup less the identity.
\end{enumerate}
\end{defn}

The following results are corollaries to the proof, rather than to the result, of Theorem~\ref{thm:BQP/RP}.

\begin{cor}\orig \label{cor:HSP}
If $\Kershort = \Eqshort$ and a language $L \in \cc{NP}$ has groupy witnesses in a family $\mathcal{G}$ of groups, then $L$ Cook-reduces to the hidden subgroup problem for the family $\mathcal{G}$.  Briefly: $L \leq_{T}^{P} \lang{HSP}(\mathcal{G})$.
\end{cor}

\begin{proof}
Let $L \in \cc{NP}$, let $W$, $G$, $\text{dec}$, $\text{mul}$, and $\text{gen}$ be as in the definition of groupy witnesses, and let $V$ be a polynomial-time verifier for $L$ such that the witnesses accepted by $V$ on input $a$ are exactly the strings in $W(a)$.  Then the equivalence relation
\[
R_{L} = \set{((a,x),(a,y)) : x=y \text{, or } \text{dec}(a,x)=\text{dec}(a,y) \text{ and } [\text{dec}(a,x)=1 \implies V(a,xy^{-1})=1] }
\]
is in $\cc{PEq}$, since $xy^{-1}$ can be computed by the polynomial-time algorithm $\text{mul}$ guaranteed in the definition of groupy witnesses.  By hypothesis, $R_{L}$ has a complete invariant $f$.  The function $f$, the function $\text{mul}$, and the generating set $\text{gen}(a)$ are a valid instance of the hidden subgroup problem.  If $a \notin L$, then $f$ is injective, and the hidden subgroup is trivial.  If $a \in L$, then the hidden subgroup is $W(a)$.  Conversely, if the hidden subgroup is trivial, then either $a \notin L$ or the identity of the group is a witness that $a \in L$, which can be easily checked. Hence $L$ reduces to the hidden subgroup problem.
\end{proof}

\begin{cor}\orig
If $\Kershort = \Eqshort$ and the language $L$ has abelian groupy witnesses, then $L \in \cc{BQP}$. \qed
\end{cor}

\begin{lem}\orig
Every language in $\cc{UP}$ has abelian groupy witnesses. \qed
\end{lem}

\begin{open} \label{open:groupy} Are there $\cc{NP}$-complete problems with abelian groupy witnesses?  Assuming $\cc{P} \neq \cc{NP}$, are there any problems in $\cc{NP} \backslash \cc{UP}$ with abelian groupy witnesses?  \end{open}

Our definition of having groupy witnesses is similar but not identical to Arvind and Vinodchandran's definition of group-definability \cite{arvindVinodchandran}.  If a set $A \in \cc{NP}$ has abelian groupy witnesses, then in general the function $a \mapsto |G(a)|$ is in $\cc{\# P}$.  If it so happens that this function is in $\cc{FP}$, then Arvind and Vinodchandran's techniques are sufficient to show that $A$ is low for $\cc{PP}$.  This may or may not be taken as evidence that such an $A$ is unlikely to be $\cc{NP}$-complete: on the one hand, Beigel \cite{beigelPP} gives an oracle relative to which $\cc{NP}$ is \emph{not} low for $\cc{PP}$, and hence $A$ could not be $\cc{NP}$-complete.  On the other hand, Toda and Ogiwara \cite{todaOgiwara} show that $\cc{PP}\oracle{\cc{PH}} \subseteq \cc{BP} \cdot \cc{PP}$ (Tarui \cite{tarui}, independently but using similar methods, strengthens this to $\cc{ZP} \cdot \cc{PP})$.  Hence, under a derandomization assumption, $\cc{NP}$ \emph{is} in fact low for $\cc{PP}$, and so the lowness of $A$ for $\cc{PP}$ is no obstruction to its being $\cc{NP}$-complete.

However, even if $|G(a)|$ is computable in polynomial time, it may yet be possible to use Corollary~\ref{cor:HSP} to show that $\Kershort = \Eqshort \implies \cc{NP} \subseteq \cc{BQP}$, as there are several classes of non-abelian, and even non-solvable, groups for which the HSP is known to be in $\cc{BQP}$ (see, \eg, \cite{grigniHSP, friedlHSP, ivanyosNonabelianHSP}).

\subsection{Hardness}
\label{sec:hardProblems}

\subsubsection{Collision-free hash functions} \label{sec:collision-free} Collision-free hash functions are a useful cryptographic primitive (see, \eg, \cite{hashSurvey}).  Proposition~\ref{prop:factoring} suggests a more general connection between the collapse $\CFshort = \Kershort$ and the existence of collision-free hash functions.  

A \definedWord{collection of collision-free hash functions} is a collection of functions $\set{h_{i} : i \in I}$ for some $I \subseteq \Sigma^{*}$ where $h_{i} : \Sigma^{|i|+1} \to \Sigma^{|i|}$ are
\begin{enumerate}
\item[1.] Easily accessible: there is a probabilistic polynomial-time algorithm $G$ such that $G(1^{n}) \in \Sigma^{n} \cap I$;

\item[2.] Easy to evaluate: there is a probabilistic polynomial-time algorithm $E$ such that $E(i,w) = h_i(w)$; and

\item[3.] Collision-free: for all probabilistic polynomial-time algorithms $A$ and all polynomials $p$ there is a length $N$ such that $n > N$ implies:
\[
\Pr_{\substack{
i = G(1^{n}) \\
(x,y) = A(i)
}}[x \neq y \text{ and } h_i(x) = h_i(y)] < \frac{1}{p(n)}.
\]
\end{enumerate}

It is not known whether collections of collision-free hash functions exist, though their existence is known to follow from other cryptographic assumptions (see, \eg, \cite{damgardCollisionFree}).  Many proposed collections of collision-free hash functions, such as MD5 or SHA, can be evaluated deterministically, that is, $E \in \cc{FP}$.

\begin{prop}\orig
\label{prop:collision-free}
\collisiontext
\end{prop}

\begin{proof}
\collisionproof
\end{proof}

\subsubsection{Subgroup equality} \label{sec:subgroupEq}
The \definedWord{subgroup equality problem} is: given two subsets $\set{g_{1},\dotsc,g_{t}}$, $\set{h_{1},\dotsc,h_{s}}$ of a group $G$ determine if they generate the same subgroup.  The \definedWord{group membership problem} is: given a group $G$ and group elements $g_{1},\dotsc,g_{t},x$, determine whether or not $x \in \langle g_{1},\dotsc,g_{t} \rangle$.  A solution to the group membership problem yields a solution to the subgroup equality problem, by determining whether each $h_{i}$ lies in $\langle g_{1}, \dotsc, g_{t} \rangle$ and vice versa.  However, a solution to the group membership problem does \emph{not} obviously yield a complete invariant for the subgroup equality problem.  Thus subgroup equality problems are a potential source of candidates for problems in $\Eqshort \backslash \Kershort$.

Note that the complexity of these problems still makes sense for non-finite groups, so long as group elements can be specified by finite strings and the group operations are computable.

Fortunately or unfortunately, the subgroup equality problem for permutation groups on $\set{1,\dotsc,n}$ has a polynomial-time canonical form, via a simple modification \cite{babaiPersonalComm} of classicial techniques \cite{sims1970, sims1971,furstHopcroftLuks,knuth} (see Example~\ref{ex:subgroupEquiv} for more of the history).

\subsubsection{Boolean function congruence} \label{sec:booleanFnEq}
Two Boolean functions $f$ and $g$ are \definedWord{congruent} if the inputs to $f$ can be permuted and possibly negated to make $f$ equivalent to $g$.  If $f$ and $g$ are given by formulae $\varphi$ and $\psi$, respectively, deciding whether $\varphi$ and $\psi$ define congruent functions is Karp equivalent to $\lang{FI}$.  If $f$ and $g$ are given by their truth tables, however, Luks \cite{luksHypergraphIso} gives a polynomial-time algorithm for deciding whether or not they are congruent.  Yet no polynomial-time complete invariant for Boolean function congruence is known.  Hence function congruence may be in $\cc{PEq} \backslash \Kershort$.  

\subsubsection{Complete problems?} \label{sec:reduction} 
Equivalence problems that are $\cc{P}$-complete under $\cc{NC}$ or $\cc{L}$ reductions may lie in $\Eqshort \backslash \Kershort$ due to their inherent difficulty.  However, we currently have no reason to believe that $\cc{P}$-completeness is related to complexity classes of equivalence problems.  Towards this end, we introduce a natural notion of reduction for equivalence problems:

\begin{defn}\orig An equivalence relation $R$ \definedWord{kernel-reduces} to an equivalence relation $S$, denoted $R \leq_{ker}^{P} S$, if there is a function $f \in \cc{FP}$ such that
\[
x \sim_{R} y \iff f(x) \sim_{S} f(y).
\]
\end{defn}

Note that $R \in \Kershort$ \ifftext $R$ kernel-reduces to the relation of equality.  Also note that if $R \leq_{ker}^{P} S$ via $f$, then $R \leq_{m}^{P} S$ via $(x,y) \mapsto (f(x), f(y))$, leading to the question:

\begin{open} \label{open:kernelVsKarp}
Are kernel reduction and Karp reduction different?  Are they different on \Eqshort?  In other words, are there two equivalence relations $R$ and $S$ (in \Eqshort?) such that $R \leq_{m}^{P} S$ but $R \not\leq_{ker}^{P} S$?
\end{open}

An equivalence relation $R \in \Eqshort$ is \definedWord{$\Eqshort$-complete} if every $S \in \Eqshort$ kernel-reduces to $R$.  For any $\Eqshort$-complete $R$, $R \in \Kershort$ \ifftext $\Kershort = \Eqshort$ \ifftext the relation of equality is $\Eqshort$-complete.  

Unlike $\cc{NP}$-completeness, however, the notion of $\Eqshort$-completeness does not become trivial if $\Kershort = \Eqshort$: the relation of equality does not kernel-reduce to the trivial relation simply because equality has infinitely many equivalence classes but the trivial relation has only one.  In particular, if $\cc{P} = \cc{NP}$ then kernel reduction and Karp reduction are distinct on $\cc{PEq}$, albeit in a rather trivial way.  The question becomes more interesting if we ask for languages $R$ and $S$ in $\cc{PEq}$ \emph{of the same densities} on which kernel reduction and Karp reduction differ.

\begin{open} \label{open:PEqComplete}
Are there $\Eqshort$-complete equivalence problems?
\end{open}

\section{Oracles} \label{sec:oracles}
In order to combine the oracles from Blass and Gurevich \cite{blassGurevich1} into a single oracle, as well as construct new oracles that simultaneously separate some classes of equivalence relations and collapse others, we introduce two notions of generic oracle. Generic oracles maintain some of the key advantages of random oracles, but allow us much greater flexibility---much of the power of finite injury arguments---in their construction\footnote{Indeed, there is a notion of genericity $\mathcal{R}$ such that results regarding $\mathcal{R}$-generic oracles are completely equivalent to results regarding random oracles \cite{solovay} (see also \cite{obt}, the paragraph just prior to Section~3.2), so generic oracle constructions can be viewed as an extension of random oracle constructions.}.
For example, it is often possible to show that some property (complexity class collapse or separation) holds relative to \emph{every} generic oracle, so that it becomes much easier to construct oracles satisfying multiple properties at once. We begin with a review of generic oracle constructions; for a more in-depth discussion, see Fenner, Fortnow, Kurtz, and Li \cite{obt}.

For those not interested in the technical details of generic oracles, the main result we will need from the next section is Lemma~\ref{lem:fundGeneric}, but we have attempted to keep the technicalities to a minimum. We only use fairly restricted versions of genericity\footnote{For the initiated: rather than treat conditions in general as perfect collections of oracles, we define a condition as a partial characteristic function with finite domain. We also require a strong form of basicness: the union of any two consistent $\mathcal{G}$-conditions (union as partial characteristic functions) must also be a $\mathcal{G}$-condition.} and all the associated concepts in this paper, allowing us to greatly simplify their discussion. Much more general versions and their uses are presented in Fenner, Fortnow, Kurtz, and Li \cite{obt}.

\subsection{Preliminaries on Generic Oracles} \label{sec:generic}
Throughout this section we will use the first construction of an oracle separating $\cc{P}$ from $\cc{NP}$ \cite{bakerGillSolovay} as a canonical example. 

Many oracle constructions proceed by finite extensions: at each stage of the construction, some requirement is to be satisfied (e.g. ``the $i$-th polynomial-time machine does not accept some fixed relativizable language $L\oracle{O}$''), and we satisfy it by specifying the oracle on finitely many more strings, leaving those strings we have previously specified untouched. In this paper, a generic oracle is one built by finite extensions which also satisfies Murphy's law: ``anything which can happen will happen.'' More prosaically, a generic oracle is built by interleaving all finite extension arguments that are ``interleavable.'' In the remainder of this section we make these ideas precise. 

A \definedWord{condition} is a partial characteristic function whose domain is finite, that is, a partial function $\sigma \colon \Sigma^{*} \to \{0, 1\}$ with $\dom(\sigma)$ finite. In more general discussions of genericity, such conditions are called \definedWord{Cohen conditions}. We say that an oracle $O$ \definedWord{extends} $\sigma$ if the characteristic function of $O$ agrees with $\sigma$ on $\dom(\sigma)$. Two conditions $\sigma_{1}, \sigma_{2}$ are \definedWord{consistent} if for every $a \in \dom(\sigma_{1}) \cap \dom(\sigma_{2})$ we have $\sigma_{1}(a) = \sigma_{2}(a)$. 

Terminologically we treat a partial characteristic function as a partial oracle/set: we write $a \in \sigma$ and say ``$a$ is in $\sigma$'' if $\sigma(a) = 1$, and similarly we write $a \notin \sigma$ and ``$a$ is not in $\sigma$'' if $\sigma(a) = 0$. We are careful not to use either terminology if $a \notin \dom(\sigma)$. 

\begin{defn} \label{def:genericity} A \definedWord{notion of genericity} is a nonempty set $\mathcal{G}$ of conditions such that
\begin{enumerate}
\setcounter{enumi}{-1}
\item (branching) for all $\sigma \in \mathcal{G}$, there are at least two distinct conditions $\tau_{1}, \tau_{2} \in \mathcal{G}$ extending $\sigma$; 

\item (generic) for all $\sigma \in \mathcal{G}$ and all $a \in \Sigma^{*} \backslash \dom(\sigma)$ there is a condition $\sigma' \in \mathcal{G}$ extending $\sigma$ such that $a \in \dom(\sigma')$; and

\item (basic) if $\sigma_{1}, \sigma_{2} \in \mathcal{G}$ are consistent, then $\sigma_{1} \union \sigma_{2} \in \mathcal{G}$.
\end{enumerate}
\end{defn}

Note that the collection of all (Cohen) conditions is a notion of genericity, typically referred to as Cohen genericity. Less trivial is the notion of $\cc{UP}$-genericity.  A $\cc{UP}$ condition is a condition which has at most one string of each length, and only has strings at lengths $tower(k)$, where the $tower$ function is defined by $tower(0) = 1$ and $tower(n+1) = 2^{tower(n)}$. The collection of all $\cc{UP}$ conditions yields the notion of $\cc{UP}$-genericity.  

A $\mathcal{G}$-generic oracle is simply one built by further and further specification by $\mathcal{G}$-conditions which satisfies an additional constraint, namely, the formal version of ``Murphy's law'' which we now present.

Throughout this section we fix a logical system that is strong enough to express all the sentences we care about; for example, Peano Arithmetic with an additional unary predicate $X$, corresponding to the oracle, will suffice. If $\varphi$ is a sentence in such a system, then an oracle $O$ satisfies $\varphi$ if $\varphi$ is true upon replacing the predicate $X$ by the characteristic function for $O$. We assume, without loss of generality from the point of view of our constructions, that the logical system has only countably many sentences. 

We say that a condition $\sigma$ \definedWord{forces} the truth of a sentence $\varphi$ if $\varphi$ is true of every oracle $O$ extending $\sigma$.  For example, $\varphi$ might be the sentence 
\begin{equation} \label{eqn:BGS}
(\exists n)[M(1^{n}) = 0 \Iff (\exists x)[|x|=n \text{ and } X(x)]].
\end{equation}
The classic argument of Baker, Gill, and Solovay \cite{bakerGillSolovay} shows how to construct a Cohen condition forcing $\varphi$.  That is, we only need to specify a finite amount of the oracle to ensure that $\varphi$ is true, regardless of how we construct the rest of the oracle.

We say that a notion of genericity $\mathcal{G}$ is strong enough to force a sentence $\varphi$ if $\varphi$ can \emph{always} eventually be forced, that is, for every $\mathcal{G}$-condition $\sigma$ there is another $\mathcal{G}$-condition $\sigma'$ extending $\sigma$ such that $\sigma'$ forces $\varphi$. We say, equivalently, that $\set{\sigma \in \mathcal{G} : \sigma \text{ forces } \varphi}$ is \definedWord{dense} in $\mathcal{G}$. In fact Baker, Gill, and Solovay essentially showed that Cohen genericity is strong enough to force (\ref{eqn:BGS}).

Finally, ``Murphy's law,'' which we require of generic oracles, is that a $\mathcal{G}$-generic oracle must force every sentence $\varphi$ that $\mathcal{G}$ is strong enough to force.

\begin{defn}[Generic Oracle] Let $\mathcal{G}$ be a notion of genericity. An oracle $O$ is $\mathcal{G}$-generic if there is a consistent collection of $\mathcal{G}$-conditions $\set{\sigma_{1}, \sigma_{2}, \dotsc}$ such that $O$ extends every $\sigma_{i}$, the $\sigma_{i}$ fully specify $O$ (that is, $\bigcup_{i} \dom(\sigma_{i}) = \Sigma^{*}$), and every sentence $\varphi$ that $\mathcal{G}$ is strong enough to force is forced by some $\sigma_{i}$. \end{defn}

We see that this definition essentially captures the idea of simultaneously interleaving all constructions that ``can be interleaved,'' that is, that $\mathcal{G}$ is strong enough to force.

\begin{lem}[Existence of $\mathcal{G}$-generic oracles] \label{lem:genericExistence} For every notion of genericity $\mathcal{G}$, $\mathcal{G}$-generic oracles exist. Furthermore, the $\mathcal{G}$-generics are dense in $\mathcal{G}$, that is, for every $\mathcal{G}$-condition $\sigma$ there is a $\mathcal{G}$-generic oracle extending $\sigma$. \end{lem}

\begin{proof}
This is essentially Lemma~3.12 of Fenner, Fortnow, Kurtz, and Li \cite{obt}, and their proof goes through \emph{mutatis mutandis}, despite our restricted definitions.
\end{proof}

Putting this all together, the way we construct generic oracles in practice is captured by the following lemma:

\begin{lem} \label{lem:fundGeneric}
Let $\mathcal{G}$ be a notion of genericity and $\varphi$ a sentence. If $\mathcal{G}$ is strong enough to force $\varphi$---that is, if every $\sigma \in \mathcal{G}$ can be extended to a $\sigma' \in \mathcal{G}$ forcing $\varphi$---then \emph{every} $\mathcal{G}$-generic oracle satisfies $\varphi$.
\end{lem}

Finally, this entire discussion relativizes. When we relativize to an oracle $A$, our formal system includes a new unary predicate which is the characteristic function of $A$, in addition to the previous unary predicate $X$ corresponding to the generic oracle. We then speak of $\mathcal{G}$-generics relative to $A$.

\subsection{Oracles for $\Eqshort$, $\Kershort$, and $\CFshort$} \label{sec:oracles2}
In this section we introduce and use two new notions of genericity. A \definedWord{one-sided transitive} condition is a (Cohen) condition $\tau$ such that
\begin{enumerate}
\item (Length restriction on the $1$-side): $1\tuple{x, y} \in \tau$ implies $|x| = |y|$, and

\item (Transitivity on the $1$-side): $1\tuple{x, y} \in \tau$ and $1\tuple{y,z} \in \tau$ implies $1\tuple{x,z} \in \tau$.
\end{enumerate}
We refer to the set of strings starting with the bit $b$ as ``the $b$-side'' of an oracle or condition. Note that in a one-sided transitive condition, all we require of the $0$-side is that $\dom(\sigma)$ is finite there. It is easily verified that one-sided transitive conditions form a notion of genericity, so by Lemma~\ref{lem:genericExistence}, one-sided transitive generics exist, and furthermore Lemma~\ref{lem:fundGeneric} applies to them.

A \definedWord{$\cc{UP}$-transitive condition} is a condition $\tau$ such that
\begin{enumerate}
\item (``$\cc{UP}$'') For each length $n$, there is at most one string of length $n$ in $\sigma$;
\item (gappy) $\sigma$ is only nonempty at lengths $tower(k)$ for some $k$. The $tower$ function is defined by $tower(0) = 1$ and $tower(n) = 2^{tower(n-1)}$;
\item (length-restricted) $\tuple{x,y} \in \sigma$ implies $|x| = |y|$.
\end{enumerate}
Note that transitivity---$\tuple{x, y} \in \tau$ and $\tuple{y,z} \in \tau$ implies $\tuple{x,z} \in \tau$---follows from the $\cc{UP}$ restriction (1) and the length restriction (3). Again it is easily verified that $\cc{UP}$-transitive conditions form a notion of genericity, so $\cc{UP}$-transitive generics exist, and Lemma~\ref{lem:fundGeneric} applies to them.

\begin{thm}\orig \label{thm:oracles}
There are oracles $A$ and $B$ relative to which $\cc{P} \neq \cc{NP}$ and
\begin{gather}
\setcounter{equation}{0}
\label{eqn:tc} \ccCF{FP\oracle{A}} \neq \ccKer{FP\oracle{A}} \neq \cc{P\oracle{A} Eq}, \\
\label{eqn:fut} \ccCF{FP\oracle{B}}_{p} = \ccKer{FP\oracle{B}}_{p} \text{ and } \ccKer{FP\oracle{B}} \neq \cc{P\oracle{B} Eq}.
\end{gather}
In fact, (\ref{eqn:tc}) holds relative to any one-sided transitive generic oracle and (\ref{eqn:fut}) holds relative to $O \oplus G$ whenever $O$ is $\cc{PSPACE}$-complete and $G$ is $\cc{UP}$-transitive generic relative to $O$.
\end{thm}

We break most of the proof into three lemmas.  The proofs of Lemmas \ref{lem:Ker!=Eq} and \ref{lem:CF!=Ker} are adaptations of the proofs of Blass and Gurevich \cite{blassGurevich1} to generic oracles.  The proof of Lemma~\ref{lem:CF=Ker} is new.  

We start by restating a useful combinatorial lemma:

\begin{lem}[Blass \& Gurevich \cite{blassGurevich1} Lemma~1] \label{lem:comb} Let $G$ be a directed graph on $2k$ vertices such that the out-degree of each vertex is strictly less than $k$.  Then there are two nonadjacent vertices in $G$. \end{lem}

Lemma~\ref{lem:comb} can be proved by a simple counting argument.

For $\cc{UP}$-transitive conditions $\sigma$ (or oracles $O$) we denote by $\sim_{\sigma}$ the corresponding equivalence relation, that is, the reflexive, symmetric closure of $\set{(x,y) : \tuple{x,y} \in \sigma}$. If $\sigma$ is only a partial function, we take care to only ever write $x \sim_{\sigma} y$ if $\tuple{x,y} \in \dom(\sigma)$. For one-sided transitive conditions $\tau$, we use the same notation $\sim_{\tau}$ to denote the equivalence relation corresponding to the $1$-side, that is, the reflexive, symmetric closure of $\set{(x,y) : 1\tuple{x,y} \in \tau}$.

\begin{lem} \label{lem:Ker!=Eq} 
Relative to any one-sided transitive generic oracle or any $\cc{UP}$-transitive generic oracle, $\Kershort \neq \Eqshort$. 
\end{lem}

\begin{proof}
The proofs for the two types of genericity are essentially identical. Let $\mathcal{G}$ be ``one-sided transitive'' or ``$\cc{UP}$-transitive'' throughout. We give the proof for one-sided transitive genericity, in which all the diagonalization happens on the $1$-side; for $\cc{UP}$-transitive genericity, drop the prefixed $1$'s throughout and only add strings at lengths $n = tower(k)$ for some $k$.

For each polynomial-time oracle Turing machine $M$, let $\varphi_{M}$ denote the sentence (often called a requirement):
\[
\varphi_{M} \defeq (\exists n)[\Ker(M\oracle{X}) \neq \sim_{X} \text{ on strings of length $n$}]
\]
By Lemma~\ref{lem:fundGeneric}, it suffices to show that any $\mathcal{G}$-condition $\tau$ can be extended to a $\mathcal{G}$-condition $\tau'$ such that $\tau'$ forces $\varphi_{M}$. For then $\varphi_{M}$ will hold for every $\mathcal{G}$-generic oracle and for every $M$, separating $\Kershort$ from $\Eqshort$.

Let $M$ be a polynomial-time oracle transducer running in time $p(|x|)$. Let $\tau$ be any $\mathcal{G}$-condition. Let $\overline{\tau}$ denote the minimal (under inclusion) extension of $\tau$ to a complete characteristic function (\ie, oracle). We show how to extend $\tau$ to another $\mathcal{G}$-condition $\tau'$ that forces $\varphi_{M}$, \ie, such that $\Ker(M\oracle{O}) \neq \sim_{O}$ for any $O$ extending $\tau'$.

Let $n$ be a length such that $p(n) < 2^{n-1}$ and $\tau$ is not defined on $1\tuple{a,b}$ for any strings $a$ and $b$ of length $\geq n$. Let $\tau'$ be the extension of $\tau$ to length $p(n)$ that is equal to $\overline{\tau}$ to length $p(n)$. If there are distinct strings $x$ and $y$ of length $n$ such that $M^{\overline{\tau}}(x)=M^{\overline{\tau}}(y)$, then $x \not\sim_{\tau'} y$ but $M^{\tau'}(x) = M^{\tau'}(y)$, and this clearly holds for any $O$ extending $\tau'$.

Otherwise, $M^{\overline{\tau}}(x) \neq M^{\overline{\tau}}(y)$ for every two distinct strings $x$ and $y$.  Say that $x$ \definedWord{affects} $y$ if $M$ queries $\overline{\tau}$ about $1\tuple{x,y}$ or $1\tuple{y,x}$ in the computation of $M^{\overline{\tau}}(y)$.  Let $G$ be a digraph on the strings of length $n$, in which there is a directed edge from $y$ to $x$ if $x$ affects $y$.  The out-degree of each vertex is at most $p(n)$, which is strictly less than $2^{n-1}$ by the choice of $n$.  Since there are $2^{n}$ vertices, Lemma~\ref{lem:comb} implies that there are two strings $x$ and $y$ of length $n$ such that neither affects the other.  Put $1\tuple{x,y}$ into $\tau'$.  Then $M^{\tau'}(x) \neq M^{\tau'}(y)$ but $x \sim_{\tau'} y$, and this holds for any oracle $O$ extending $\tau'$.

Thus $\Kershort\oracle{O} \neq \Eqshort\oracle{O}$ relative to any $\mathcal{G}$-generic oracle $O$, for $\mathcal{G}$ either ``one-sided transitive'' or ``$\cc{UP}$-transitive.''
\end{proof}

\begin{lem} \label{lem:CF!=Ker} 
Relative to any one-sided transitive generic oracle, $\CFshort \neq \Kershort$.
\end{lem}

\begin{proof}
For this proof, all the diagonalization is performed on the $0$-side.

We describe our oracles $O$ and conditions $\tau$ with values in the alphabet $\set{0,1,2}$ for simplicity (that is, $\tau\colon \Sigma^{*} \to \set{0, 1, 2}$).  Let $read\oracle{O}\colon \Sigma^{*} \to \Sigma^{*}$ denote the oracle function
\[
read\oracle{O}(x) = O(0x01)O(0x011) \cdots O(0x01^{k-1})
\]
where $k$ is the least value such that $O(0x01^k)=2$.  Note that the bits used by $read\oracle{O}$ on input $x$ are disjoint from those used by $read\oracle{O}$ on any input $y \neq x$. Also note that $read\oracle{O}$ only queries the oracle regarding strings on the $0$-side. Let $R\oracle{O} = \Ker(read\oracle{O})$. 

Let $f$ be any polynomial-time oracle transducer, and define
\[
\psi_{f} \defeq (\exists n)[f\oracle{X} \text{ is not a canonical form for } R\oracle{X} \text{ on strings of length $n$}].
\]
As in Lemma~\ref{lem:Ker!=Eq}, it suffices to show that any one-sided transitive condition $\tau$ can be extended to a one-sided transitive condition $\tau'$ forcing $\psi_{f}$, by Lemma~\ref{lem:fundGeneric}.

Let $f$ be a polynomial-time oracle transducer running in time $p(|x|)$. Let $\tau$ be a one-sided transitive condition, and let $\overline{\tau}$ denote the oracle extending $\tau$ which has value $2$ on strings of the form $0x$ that are not in $\dom(\tau)$ and value $0$ on all other strings not in $\dom(\tau)$. We show how to extend $\tau$ to a one-sided transitive condition $\tau'$ such that $f\oracle{O}$ does not compute a canonical form for $R\oracle{O}$ for any $O$ extending $\tau'$.

Let $n$ be a length such that $p(n) < 2^{n-1}$ and such that $\tau$ is not defined for any strings $0x$ with $|x| \geq n$. For a string $x$ of length $n$, let $\tau_{x}$ denote the minimal extension of $\tau$ such that $read\oracle{\overline{\tau_{x}}}$ is the identity on all strings of length $n$, except $read\oracle{\overline{\tau_{x}}}(x)=1^{n+1}$. Since the $read$ function only queries strings on the $0$-side, $\tau_{x}$ differs from $\tau$ only on the $0$-side, and we do not need to worry about violating transitivity on the $1$-side.  Note that $read\oracle{\overline{\tau_{x}}}$ is injective on strings of length $n$, so its kernel at length $n$ is the relation of equality.  In particular, any canonical form for $R\oracle{\overline{\tau_{x}}}$ must be the identity on strings of length $n$.

If there is an $x$ of length $n$ such that $f\oracle{\overline{\tau_{x}}}(x) \neq x$, then $f\oracle{\overline{\tau_{x}}}(x)$ is not the identity on strings of length $n$, so $f\oracle{\overline{\tau_{x}}}$ is not a canonical form for $R\oracle{\overline{\tau_{x}}}$.  Let the extension $\tau'$ be $\overline{\tau_{x}}$ up to length $p(n)$.

Otherwise, $f\oracle{\overline{\tau_{x}}}(x)=x$ for all $x$ of length $n$. We say that $f\oracle{O}(x)$ queries the oracle about $y$ if $f\oracle{O}(x)$ queries any of the strings that $read\oracle{O}(y)$ queries. Find $x$ and $y$ of length $n$ such that $f\oracle{\overline{\tau_{x}}}(x)$ does not query the oracle about $y$ and $f\oracle{\overline{\tau_{y}}}(y)$ does not query the oracle about $x$.  This is possible by Lemma~\ref{lem:comb}, as in the proof of Lemma~\ref{lem:Ker!=Eq}. Let $\tau'$ be the minimal oracle extending $\tau$ such that $read\oracle{\tau'}$ is the identity on strings of length $n$, except $read\oracle{\tau'}(x) = read\oracle{\tau'}(y) = 1^{n+1}$.  Then $\tau'$ differs from $\overline{\tau_{x}}$ only on those strings in its domain queried by $read\oracle{\tau'}(y)$ and $\tau'$ differs from $\overline{\tau_{y}}$ only on those strings in its domain queried by $read\oracle{\tau'}(x)$.  Since $f\oracle{\overline{\tau_{x}}}(x)$ does not query the oracle about $y$ we have $f\oracle{\overline{\tau_{x}}}(x) = f\oracle{\tau'}(x) = x$ and similarly $f\oracle{\overline{\tau_{y}}}(y) = f\oracle{\tau'}(y) = y$. So relative to any oracle $O$ extending $\tau'$, we have $(x,y) \notin \Ker(f\oracle{O})$ but  $read\oracle{O}(x)=read\oracle{O}(y)=1^{n+1}$. Again, $\tau'$ forces that $f\oracle{\tau'}$ is not a canonical form for $R\oracle{\tau'}$.

Thus $\CFshort\oracle{O} \neq \Kershort\oracle{O}$ relative to any one-sided transitive generic oracle $O$. 
\end{proof}

\begin{lem}\orig \label{lem:CF=Ker} 
If $\cc{P}=\cc{PSPACE}$, and $O$ has at most one string of each length $tower(k)$ and no other strings, then $\ccCF{FP\oracle{O}}_{p} = \ccKer{FP\oracle{O}}_{p}$. Furthermore, this result relativizes. \end{lem}

\begin{proof} Let $O$ have at most one string of each length $tower(k)$, and no other strings.  Let $f$ be an oracle transducer running in polynomial time $p(|x|)$, let $R = \Ker(f\oracle{O})$, and suppose that $\tuple{x,y} \in R$ implies $|x| \leq q(|y|)$ for some polynomial $q$.  For any input $x$ of sufficient length, all elements of $O$ except possibly one have length either $\leq \log p(|x|)$, in which case they can be found rapidly, or $> p(q(|x|))$ in which case they cannot be queried by $f$ on any input $y \sim_{R} x$.  Following a technique used in \cite{twoQueries1}, we call this one element the ``cookie'' for this equivalence class.

For the remainder of this proof, ``minimum,'' ``least,'' etc. will be taken with respect to the standard length-lexicographic ordering. 

We show how to efficiently compute a canonical form for $R$.  Let $R_{y}$ denote the inverse image of $y$ under $f\oracle{O}$, which is an $R$-equivalence class.  Let
\begin{eqnarray*}
B_{y} & = & \set{x : f\oracle{O}(x) = y \text{ and } f\oracle{O}(x) \text{ does not query the cookie}},
\end{eqnarray*}
$r_{y} = \min R_{y}$, and $b_{y} = \min B_{y}$.  A canonical form for $R$ is
\[
g(x) = \begin{cases}
b_{y} & \If B_{y} \neq \emptyset  \\
r_{y} & \otherwise,
\end{cases}
\]
where $y=f\oracle{O}(x)$.  Now we show that $g$ is in fact in $\cc{FP\oracle{O}}$.  On input $x$, the computation of $g$ proceeds as follows:
\begin{enumerate}
\item\label{item:FfindSmall} Find all elements of $O$ of length at most $\log p(|x|)$.  Any further queries to $O$ of length $\leq \log p(|x|)$ will be simulated without queries by using this data.

\item Compute $y=f\oracle{O}(x)$.  

\item If the cookie was queried, then \emph{all} further queries to $O$ will be simulated without queries using this data.  Using the power of $\cc{PSPACE}$, determine whether or not $B_{y} = \emptyset$.  If $B_{y} = \emptyset$, find and output $r_{y}$.  If $B_{y} \neq \emptyset$, find and output $b_{y}$.

\item If the cookie was not queried, then $x \in B_{y}$, so $B_{y} \neq \emptyset$.  Use the power of $\cc{PSPACE}$ to find the least $z$ such that $f(z) = y$, answering $0$ to any queries made by $f$ to strings of length $\ell$ between $\log p(|x|) < \ell \leq p(q(|x|))$.

\item Run $f\oracle{O}(z)$.  If $f\oracle{O}(z)$ did not query the cookie, then $f\oracle{O}(z)=f(z) = y$ and $z = b_{y}$, so output $z$.  Otherwise, $f\oracle{O}(z)$ queried the cookie, so no further oracle queries need be made.  Using the power of $\cc{PSPACE}$, find and output $b_{y}$.
\end{enumerate}
\end{proof}

\begin{proof}[Proof of Theorem~\ref{thm:oracles}]
($\CFshort \neq \Kershort \neq \Eqshort$) By Lemmas~\ref{lem:Ker!=Eq} and \ref{lem:CF!=Ker}, $\CFshort \neq \Kershort \neq \Eqshort$ relative to any one-sided transitive generic oracle.

($\cc{\CFshort_{p}} = \cc{\Kershort_{p}}$ and $\Kershort \neq \Eqshort$) Relativize to any $\cc{PSPACE}$-complete set $C$, let $O$ be any $\cc{UP}$-transitive generic oracle relative to $C$, and \emph{re}relativize to $O$. Note that Lemma~\ref{lem:Ker!=Eq} relativizes, so relative to $C$ and $O$ combined, $\Kershort \neq \Eqshort$. Since $\cc{P} = \cc{PSPACE}$ relative to $C$, and $O$ has at most one string of each length $tower(k)$ and no other strings, and Lemma~\ref{lem:CF=Ker} relativizes, we also have $\CFshort_{p} = \Kershort_{p}$ relative to $C$ and $O$ combined.
\end{proof}

\begin{open} \label{open:CFvsKer} Does $\CFshort = \Kershort$ imply $\cc{P} = \cc{NP}$?  Or is there an oracle relative to which $\CFshort = \Kershort$ but nonetheless $\cc{P} \neq \cc{NP}$?  Further, is there an oracle relative to which $\cc{P} \neq \cc{NP}$ but $\CFshort = \Kershort = \Eqshort$?  \end{open}

\begin{open} \label{open:CF!=Ker=PEq} Is there an oracle relative to which $\CFshort \neq \Kershort = \Eqshort$? \end{open}

\section{Future Work}
Here we present several directions for future work, in addition to the open problems mentioned throughout the paper.  

\subsection{Logarithmic Space}
It would also be interesting to study equivalence relations decidable in logarithmic space. 

For example, it has been shown that the word equality problem (given two words in the generators of a group, do they represent the same group element?) for a finitely generated linear group is decidable in logarithmic space \cite{liptonZalcstein, simonWordProblem}.  (A group is linear if it is isomorphic to a group of matrices over some field.)  In fact, implicit in the proofs is a log-space complete invariant: essentially the matrix corresponding to a word in the generators. But it seems unlikely that, in general, one can get from the matrix a corresponding canonical form, that is, a canonical word in the group generators representing each group element. Hence the word problem in finitely generated linear groups is a potential witness to $\ccKer{FL} \neq \ccCF{FL}$. One open problem is to explicitly construct a linear group with no log-space canonical form for its word equality problem.

Analogues of many of the results in this paper for logarithmic space are intriguing open questions: 

\begin{itemize}
\item Is $\cc{LEq}$ contained in $\ccCF{FL\oracle{\cc{NL}}}$?  Is it contained in $\ccCF{FP}$?  In $\ccKer{FP}$?  We note that the straightforward binary search technique used to show $\cc{PEq} \subseteq \ccLexEq{FP\oracle{\cc{NP}}}$ does not work in logarithmic space.  Jenner and Tor\'{a}n \cite{jennerToran} showed that the lexicographically minimal (or maximal---in this case the same technique works) solution of any $\cc{NL}$ search problem can be computed in $\cc{FL}\oracle{\cc{NL}}$.  However, the notion of an $\cc{NL}$ search problem is based on the following characterization of $\cc{NL}$ due to Lange \cite{langeNL}: a language $A$ is in $\cc{NL}$ \ifftext there is a a polynomial $p$ and a log-space machine $M(x,\vec{y})$ that reads its second input in one direction only, indicated by ``$\vec{y}$'', such that
\[
x \in A \iff (\exists y : |y| \leq p(|x|))[M(x,\vec{y})=1].
\]
Without the one-way restriction, this definition would give a characterization of $\cc{NP}$ rather than $\cc{NL}$.  An $\cc{NL}$ search problem is then: given such a machine $M$ and input $x$, find a $y$ such that $M(x,\vec{y})=1$.  Any equivalence relation that can be decided by such a machine---that is, where $x \sim y$ \ifftext $M(x, \vec{y}) = 1$---is in $\ccLexEq{FL\oracle{\cc{NL}}}$, but it is not clear that this captures all of $\cc{LEq}$.

\item Does $\ccCF{FL} = \ccKer{FL}$ imply $\cc{NL} = \cc{UL}$?  Note that $\cc{NL} = \cc{UL}$ \ifftext $\cc{FL}\oracle{\cc{NL}} \subseteq \cc{\# L}$ \cite{alvarezJennerSharpL}.

\item Does $\ccCF{FL} = \cc{LEq}$ imply $\cc{UL} \subseteq \cc{RL}$?  A positive answer to this question and the previous one would give very strong evidence that $\ccCF{FL} \neq \cc{LEq}$, as significant progress has been made towards showing $\cc{L} = \cc{RL}$ \cite{reingoldTrevisanVadhan}.
\end{itemize}

\subsection{Additional Questions}
In no particular order:
\begin{itemize}
\item In Example~\ref{ex:formulaEquiv} we observed that Boolean formula equivalence is a natural equivalence relation that is $\cc{coNP}$-complete. The equivalence relation generated by $0x \sim_{R} 1x$ if and only if $x \in \lang{SAT}$ is clearly $\cc{NP}$-complete, but is not particularly natural as an equivalence relation.  Are there \emph{natural} $\cc{NP}$-complete equivalence relations?

\item Study expected polynomial-time canonical forms.  If every $R \in \ccKer{FP}$ has an expected polynomial-time canonical form, does $\cc{PH}$ collapse? An interesting example of an expected polynomial-time canonical form is that for graph isomorphism \cite{babaiKucera}.

\item Find a class of groups for which the group membership problem is in $\cc{P}$ but no efficient complete invariant is known for the subgroup equality problem (see Section \ref{sec:subgroupEq}).

\item If $\Kershort = \Eqshort$, does $\cc{PH}$ collapse?

\item $\ccLexEq{FP\oracle{\cc{\Sigma_{i} P}}} \stackrel{?}{=} \ccCF{FP\oracle{\cc{\Sigma_{i} P}}} \stackrel{?}{=} \ccKer{FP\oracle{\cc{\Sigma_i P}}} \stackrel{?}{=} \cc{P\oracle{\cc{\Sigma_i P}} Eq}$.  If $\ccKer{FP\oracle{\cc{\Sigma_{i} P}}} = \cc{P\oracle{\cc{\Sigma_{i} P}} Eq}$ does $\cc{PH}$ collapse?

\item Study counting classes of equivalence relations.  For an equivalence relation $R$, the associated counting function is $f(x) = \#\set{y : y \sim_{R} x}$.

\item Preorders have been studied in the context of $p$-selectivity and semifeasible sets \cite{ko83}, and partial orders have been studied in the context of $\cc{\# P}$ and acceptance mechanisms for nondeterministic machines \cite{partialOrders}. It would be interesting to develop these further, as well as to study complexity classes of lattices and total orders. 

\end{itemize}

\section*{Acknowledgments}
The authors thank Stuart Kurtz and Laci Babai for several useful discussions.  In particular, Stuart suggested the use of the equivalence relation $R_{L}$, which led us to Theorem \ref{thm:BQP/RP}, and Laci pointed out the canonical form for subgroup equality of permutation groups \cite{babaiPersonalComm}.  We thank Scott Aaronson for the observations leading to Section \ref{sec:promise}.  We thank Andreas Blass for pointing us to the original two papers he co-authored with Gurevich \cite{blassGurevich1, blassGurevich2}. We thank Paolo Codenotti for useful comments on a draft. Finally, we thank the editor, Lane Hemaspaandra, and two anonymous reviewers for suggestions that significantly improved the clarity and the organization of the paper. In particular, one of the reviewers suggested that we define some sort of hybrid notion of Cohen and transitive genericity, as well as suggested the notion of $\cc{UP}$-transitive genericity.

\bibliographystyle{ams-alph}
\bibliography{kernels}

\end{document}